\documentclass[onecolumn]{aastex631}
\usepackage{amsmath}
\usepackage{multirow} 

\begin{document}

\title{ODNet: A Convolutional Neural Network for Asteroid Occultation Detection}

\author[0000-0002-7123-1813]{Dorian Cazeneuve}
\affiliation{SETI Institute \\
339 Bernardo Ave, Suite 200\\
Mountain View, California, 94043, USA}
\affiliation{Unistellar \\
5 All. Marcel Leclerc, batiment B\\
Marseille, 13008, France}

\author[0000-0001-7016-7277]{Franck Marchis}
\affiliation{SETI Institute \\
339 Bernardo Ave, Suite 200\\
Mountain View, California, 94043, USA}
\affiliation{Unistellar \\
5 All. Marcel Leclerc, batiment B\\
Marseille, 13008, France}

\author[ 0000-0002-0973-4276]{Guillaume Blaclard}
\affiliation{Unistellar \\
5 All. Marcel Leclerc, batiment B\\
Marseille, 13008, France}

\author[0000-0002-4297-5506]{Paul A.\ Dalba}
\altaffiliation{Heising-Simons 51 Pegasi b Postdoctoral Fellow}
\affiliation{SETI Institute \\
339 Bernardo Ave, Suite 200\\
Mountain View, California, 94043, USA}
\affiliation{Department of Astronomy and Astrophysics, University of California \\
Santa Cruz, CA 95064, USA}

\author[0000-0002-8619-6897]{Victor Martin}
\affiliation{Unistellar \\
5 All. Marcel Leclerc, batiment B\\
Marseille, 13008, France}

\author{Joe Asencio}
\affiliation{Unistellar \\
5 All. Marcel Leclerc, batiment B\\
Marseille, 13008, France}

\begin{abstract}

We propose to design and build an algorithm that will use a Convolutional Neural Network (CNN) and observations from the Unistellar network to reliably detect asteroid occultations. The Unistellar Network, made of more than 10,000 digital telescopes owned by citizen scientists, and is regularly used to record asteroid occultations. In order to process the increasing amount of observational produced by this network, we need a quick and reliable way to analyze occultations. In an effort to solve this problem, we trained a CNN with artificial images of stars with twenty different types of photometric signals. Inputs to the network consists of two stacks of snippet images of stars, one around the star that is supposed to be occulted and a reference star used for comparison. We need the reference star to distinguish between a true occultation and artefacts introduced by poor atmospheric condition. Our Occultation Detection Neural Network (ODNet), can analyze three sequence of stars per second with 91\% of precision and 87\% of recall. The algorithm is sufficiently fast and robust so we can envision incorporating onboard the eVscopes to deliver real-time results. We conclude that citizen science represents an important opportunity for the future studies and discoveries in the occultations, and that application of artificial intelligence will permit us to to take better advantage of the ever-growing quantity of data to categorize asteroids.

\end{abstract}


\section{Introduction} \label{sec:intro}
The Unistellar network is a worldwide network of citizen scientists and professional astronomers who collaborate on a daily basis to scientific campaigns. These citizen scientists are equipped with one of the Unistellar's digital telescopes: the eVscope 1 \citep{MARCHIS202023} , the eQuinox (same architecture as the eVscope 1) or the eVscope 2.  Those three models have redundancies and differences detailed in the Table \ref{table:evscopes}. \\ 

\begin{table}[t]
\caption{Hardware Comparison of eVscopes} 
\label{table:evscopes} 
\centering    
\begin{tabular}{c c c}
     \hline\hline
     Characteristics & eVscope 1 & eVscope 2 \\
    \hline
     Pixels per Image & $1\,304 \times 996$ & $2\,048 \times 1\,536$ \\
     Sensor& IMX224 & IMX347\\
     Aperture ($mm$) & \multicolumn{2}{c}{114.3} \\
     Focal Length ($mm$) & \multicolumn{2}{c}{450}\\
     Field of View ($arcsec^2$) & $37 \times 27 $  & $47 \times 34 $ \\
     Pixel Scale ($arcsec\cdot pixel^{-1}$) & 1.72 & 1.33\\
    \hline
\end{tabular}
\end{table}

Because of their hardware similarities (mirror, aperture, focal length, sensors, pixel scale and sensitivity), researchers at the SETI Institute, Unistellar's scientific partner, can receive standardized data sets for curated scientific campaigns. Those campaigns are often organized by SETI institute researchers, and announced on the Unistellar's \href{https://unistellaroptics.com/citizen-science/}{website} . Professional  astronomers  can also  request  observations to the Unistellar network. For selected campaigns, such as asteroid occultations, exoplanet transits, comets, or near-Earth asteroid (NEA) observations, citizen scientists can learn how to get involved on the Unistellar’s website. The information available there includes the timing of the observations, the locations to be observed, the celestial coordinates, and the observing parameters (exposure, gain, duration) to be used on the eVscope.  

One of the most popular Unistellar scientific programs is related to occultations by asteroids, which are astronomical events defined as when an asteroid passes between a star and an observer located on Earth, hiding the star for a brief moment. From the observer’s point of view, the star will disappear from less than a second up to more than a minute depending on the size, relative velocity and position of the asteroid. Recording such events helps astronomers retrieve the characteristics of asteroids including its astrometric positions, an estimate of its projected shape \citep{braga2014ring} and even whether moons or rings surround it \citep{gibney2014asteroids}.

The Gaia mission now offers accurate astrometric measurements of bright stars (typically $G<18$) which allows astronomers to increase the accuracy of occultation predictions, and maximize the chance of successful observations. Over the past forty years, occultation have been used to derive the size, shape, and multiplicity of asteroids. Combined with other techniques such as radar detection (\cite{ostro2000radar}), direct imaging by adaptive optics on ground-based telescopes (\cite{descamps2007figure} and \cite{vernazza2021vlt}) or using the Hubble Space Telescope \citep{parker2006ceres}), astronomers can gain new insights into asteroids. 
Unfortunately because only a handful number of space missions have flown past asteroids \citep{barucci2007rosetta}, and even fewer have orbited them \citep{russell2011dawn}, today our knowledge about these asteroids relies mostly on remote observing methods. Because  occultation events are numerous and can be performed for any type of asteroids (from the NEA population to the distant Trans-Neptunian objects), they are probably the most promising techniques we have to characterize asteroids in coming years.

Occultation combined with light-curve inversion (\cite{vdurech2011combining}, \cite{viikinkoski2015adam}) is a powerful way to estimate the size of many asteroids. To date, the DAMIT \citep{vdurech2010damit} database contains the size estimate of 3\,462 asteroids most of them in the main-belt and with a diameter greater than 10 km.
Determining the size and shape of asteroids is one key to understanding their formation and evolution. For instance,  the elongated and bilobated shape of the asteroid (216) Kleopatra is linked to its critically rotating state, which probably formed the moons \citep{marchis2021216}. \newline
One caveats of the occultation technique is the need for a large number of observers able to properly estimate an object's shape and size. Highly sampled occultation are rare, since they require a large number of observers under the path (which has a width similar to the projected size of the asteroid). The odds that a prediction is accurate enough to detect the occultation is high for those observers. One of the most densely viewed occultation to date was a campaign organized with the professional and amateur community in California targeting the binary asteroid (90) Antiope \citep{colas2012}. \newline
A classic occultation lasts from less than a second to twenty seconds, while the total duration of an observation might be five to fifteen minutes long due to uncertainties. The exposure time (i.e., the time between the opening and closing of the shutter while the detector is continuously acquiring photons) is then set to 200-300 milliseconds, meaning that an observation for one user can represent as much as 3\,000 frames. Note that the occultation appears only on $\sim$60 of those frames in the best cases (and 5 to 20 frames in most of the time). Given the number of observers for each event the amount of information to treat for scientific teams can swiftly be overwhelming. \\

As the Unistellar community grows, so does the amount of data acquired in asteroid occultations. This means that the rapid manual identification of occultations and subsequent processing can become unsustainable. 
This can lead to a bottleneck in refining the asteroid ephemerides and characteristics. 
The slowdown may also reduce citizen scientist interest in observing asteroid occultations. This issue is shared by any large network of observers conducting asteroid occultation observations. Also, the legacy detection code, that will be described in Section \ref{section:2} yields a critical number of false results, increasing delays in data analysis and requiring additional human involvement. We clearly need for an accurate, fast, and automated tool to confirm or refute the presence of an occultation in a photometric data set. 
To solve this problem, the SETI-Unistellar team worked to develop a faster but nonetheless reliable solution based on machine learning to detect occultation as seen from an eVscope. Because we know exactly which star should disappear as we predict the occultation, we can reduce the area of interest to the pixels of that star and its neighbourhood. \\ 

Between March 2020 and June 2022,  Unistellar collected $\sim$1014 occultation observations, with 132 positive and 394 negative result. The remaining observations were unusable due to a bad pointing, weather issues and other environmental problems during the observation. This database is always growing : it represented 217 individual observation on the year 2020, 461 observations in 2021, and 336 in the first semester of 2022.
\newline

%
The articleis organized as follows. In Section 2, we examine the occultation method currently used by the science team. In Section 3, we conduct a global modeling of the CNN model, named ``ODNet'' (Occultation Detection Neural Network), justifying our choices of architecture and the application's expected inputs and outputs for the application. In Section 4, we confirm the efficiency of the trained network, testing it on real data and using different metrics to determine if this new method is more efficient than the legacy code. In Section 5, we discuss the characteristics of our solution for the automatic detection of occultation. In Section 6, we summarize our findings and explain their potential for future works they imply.

\section{Standard Detection of an Occultation with an eVscope}\label{section:2}

Let us now describe the analysis currently in use in the SETI-Unistellar science team for the detection of occultation on the eVscope network. From here, we will refer to the following method as the ``legacy code'' in opposition to the ML method.

For purpose of this study, we assume that all frames captured by an eVscope are saved as .FITS files \citep{wells1979fits} and have size $1305 \times 997$ pixels$^2$ or $2048 \times 1536$ pixels$^2$ depending on the eVscope model. To create a uniform data set, eVscope 2 frames are reduced to eVscope 1 format, making a typical star selected for an observation five to twenty-five pixels wide, depending on its magnitude (usually between $V$=7 and $V$=13).

We apply a two-dimensional Gaussian filter to the raw frame to soften the point spread function (PSF) (i.e., improve the roundness of the shape of the stars) and to smooth over some of the background noise. 

\begin{figure}[!ht]
   \centering
   \includegraphics[width=8.5cm]{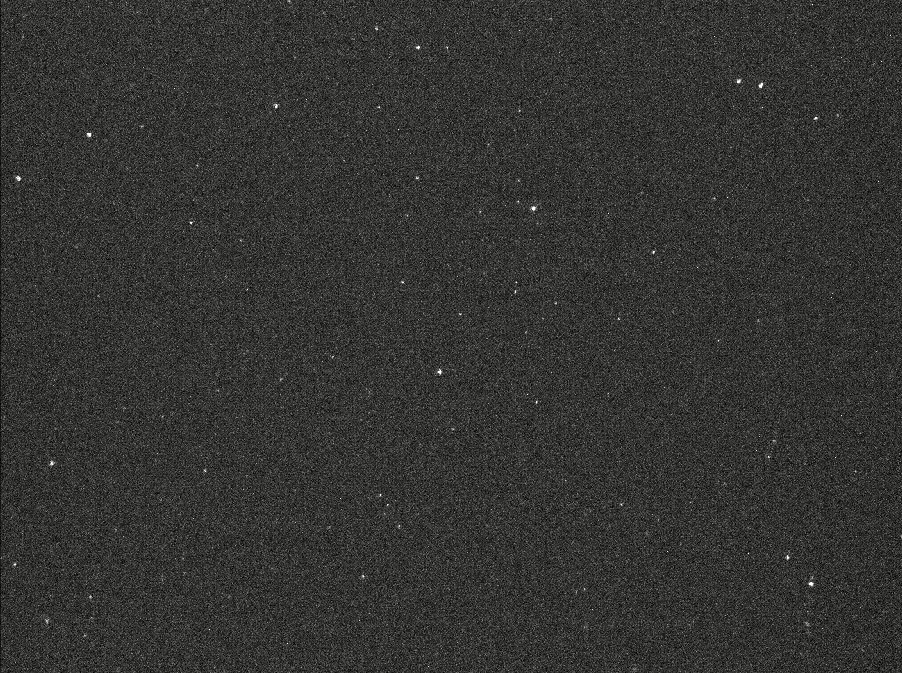}
   \includegraphics[width=8.5cm]{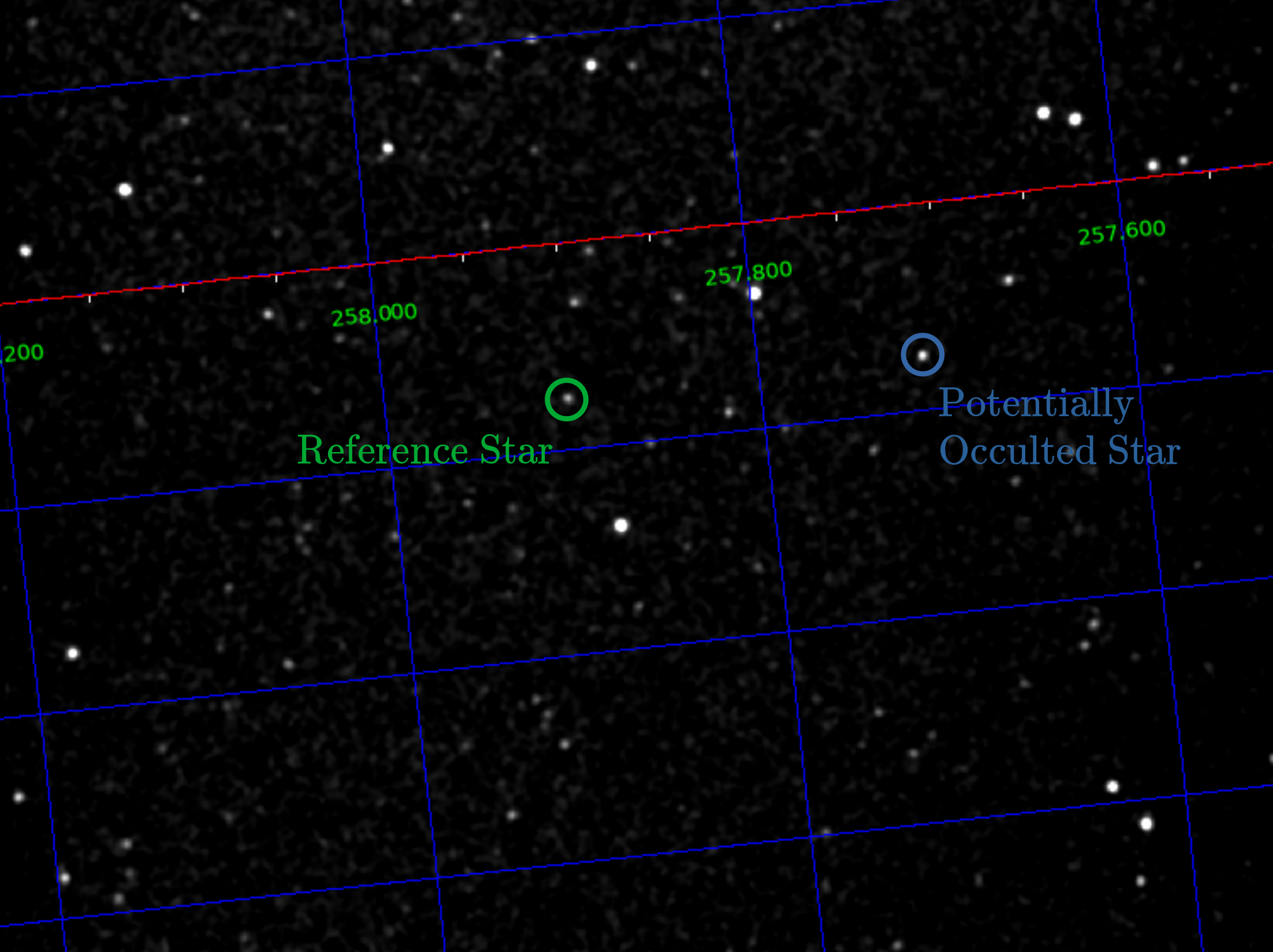}
      \caption{Visualization of the process from a raw picture to a Gaussian-blurred frame to an AFD-identified frame with a coordinate grid of the RaDec in degree and the stars selected for study.
      From left to right: 1-Raw frame 2-Blurred and Post AFD frame. The display for the left image was adjusted so the low level signal, the faint stars and background noise, is visible.
              }
         \label{FigProMisa}
\end{figure}

Also, we subtract a dark frame and the background if needed. Some observers are located in urban areas, having Bortle scale of 5 or 6, making ambient light pollution prominent).  \newline
Next, we conduct an autonomous field detection (AFD), which is effectively an optimized plate solution, and we define a reference star to use as a comparison for the possibly occulted target star. We conduct an aperture photometry of the target and reference stars via \texttt{Photutils} \citep{larry_bradley_2022_6385735} to determine their flux, flux error and signal-to-noise ratio (S/N) as a function of time. The transition from raw to a reduced, plate-solved data is illustrated in Figure \ref{FigProMisa}.

\begin{figure}[!ht]
   \centering
   \includegraphics[width=12cm]{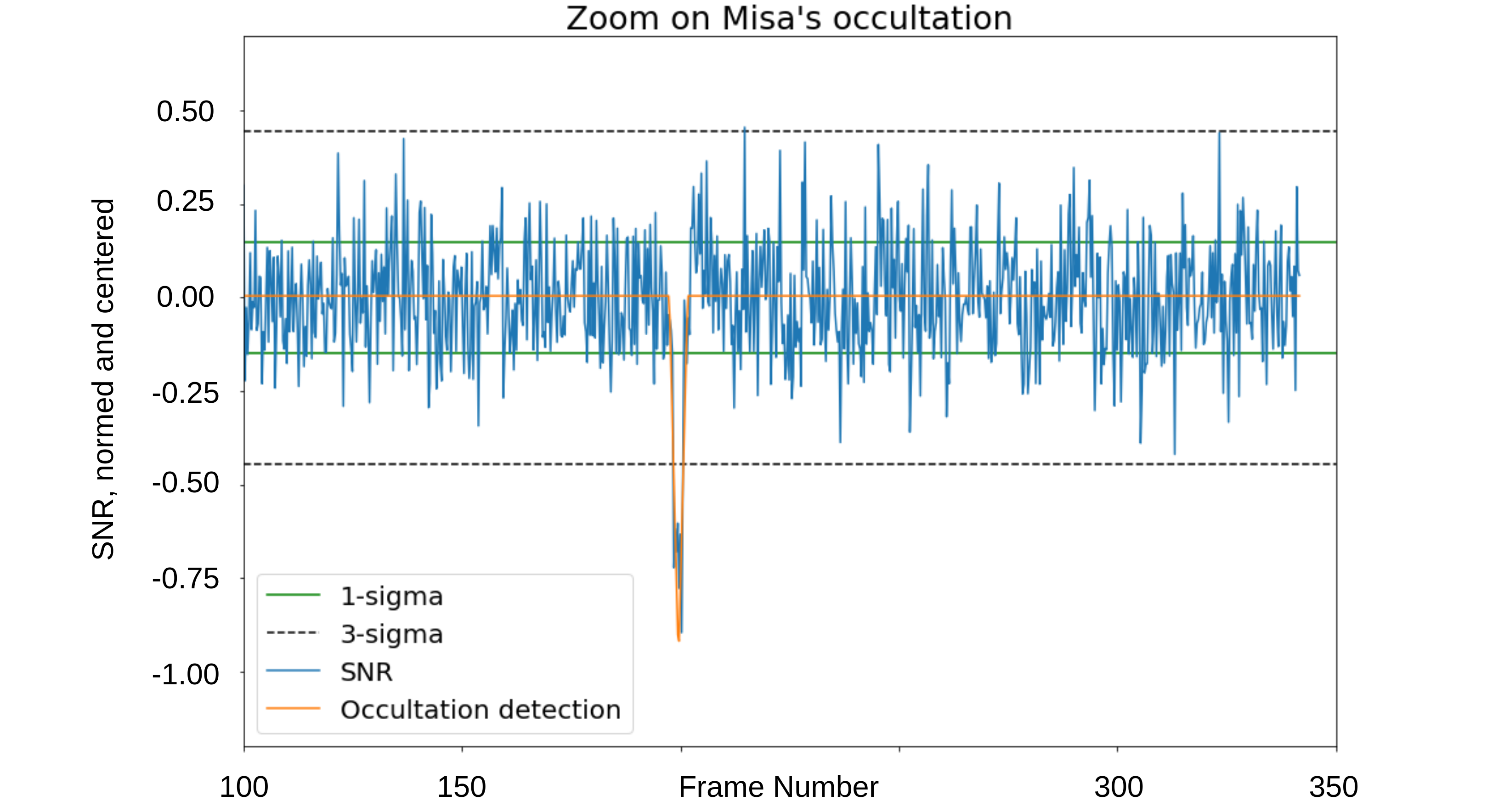}
      \caption{Detection of an occultation using the legacy method.
              }
         \label{FigPhotoMisa}
\end{figure}

\text{After completing the photometry}, we try to fit a rectangular window function to the S/N and flux time series in order to identify an occultation as described in \cite{2020SciPy-NMeth}. If this code identifies an occultation in both flux and S/N, the rectangular function's parameters provide us with the timing of the disappearance, reappearance, and incertitude of the observation. As seen in Figure \ref{FigPhotoMisa}, the beginning and end of the rectangular window function correspond to the disappearance and reappearance of the star, respectively. In this figure, the S/N has been  normalized, the saturated point erased, and the function for occultation detection fitted. The resulting timing information can then be inverted to conduct studies of an asteroid's position and shape. Alternatively, if the rectangular window function cannot be fit to the flux and S/N, we declare a negative result (non-detection). 

Although the legacy code is effective for clear, high S/N occultations, it also generates many false positives detections (see Table \ref{table:metric}), because the window function can often be fit to a noise feature resulting from wind, clouds or variable atmospheric seeing. If left unchecked, the legacy code frequently finds occultation where there are none.In the next section, we will explain why we seek more efficient methods via deep-learning.

\section{Methods} \label{sec:Methods}

In this section, we will explain why we are using deep neural networks (NN), how we built our model and the tool used for this purpose.
This study relies on the library \texttt{Tensorflow 2.4.0} \citep{tensorflow2015-whitepaper} on \texttt{Python 3.7} \citep{10.5555/1593511}. 

\subsection{Deep Learning}

In the last decade, deep neural networks have shown increased performance in many classification tasks compared to traditional methods. Traditional methods compute representations of input signals using a fixed algorithm such as Principal Component Analysis (\cite{pearson_liii_1901}), Fourier or wavelet decomposition (\cite{howell_principles_2001, stephane_chapter_2009}), and then learn the highly non-linear mapping between representations and classes to predict, in our case, the class is ``presence of occultation in the presented sample''. The computation of these representations being critical, deep convolutional neural networks have been introduced to use the raw signal as input, compute an internal representation, and predict a class, all in a learn-able end-to-end framework (\cite{lecun_deep_2015}). Most of the significant advancements done the machine-learning community have been achieved in challenges using Deep Neural Networks like Face Recognition, Object Recognition, and Scene Segmentation  (\cite{russakovsky_imagenet_2015, zhu_masked_2021,deng_lightweight_2019}). Therefore, we propose to train a Deep Neural Network to estimate if an observation contains an occultation, as it represents an encouraging solution to our precision problem ,the slow response time, and the scalability of detection.

To achieve this goal, a training set is crucial for deep-learning based methods. That is the reason why we detail the characteristics of our training set in the next section. \newline

\subsection{Training Set: Simulating Stars}
Typically, an eVscope occultation observation lasts for several minutes ($\sim$1\,500 frames). Of these frames, only $\sim$10 contain the occultation. At the onset of this study (March 2021), the entire Unistellar database only contained $\sim$100 usable occultation data sets. About 15\% of these were positive, meaning that we had only $\sim$150 frames containing an occultation.
Owing to the modest size of this data set and the fact that we cannot use the same data for training and testing our model, we decided to synthesize artificial data to use as the training set. Later, we will apply the trained model to the real positive occultation data to test its abilities.

In generating a synthetic training set, we wanted to explore all possible photometric variation scenarios for occultations with the goal of making our neural network as robust as possible. 
To simulate stars in synthetic eVscope data, we inspected every eVscope frame collected by the entire network between October 2020 and March 2021. 
From these frames, we measured the statistical distributions of stellar fluxes and background values. We then synthesized star snippets from these distributions and blurred the snippets with a Gaussian filter. We noticed that stars, when blurred, represented a more constant and reproducible pattern, showing a Gaussian distribution of light around the star, and a noisy background as it can be seen in Figure \ref{FigCompareSet}. Therefore, it represents a more consistent way to produce artificial data in large quantity and variety.

In order to make the most reliable simulation of a star seen by eVscopes and usable as an input for the deep-learning application, we must first normalize every set of snippet images of one star (stack) so that the intensity spans from 0 to 1. Specifically, we determine the maximum pixel value in a stacks of N images and scale all of the snippets in that stack such that the maximum is 1 and the background is near 0. We do this for the N-image stack of the target and reference star separately, so that each one is scaled between 0 and 1 even if one star is far brighter than the other. In this way, the scaled intensity of the target and reference star are very similar.

\begin{figure*}
   \centering
   \includegraphics[width=9cm]{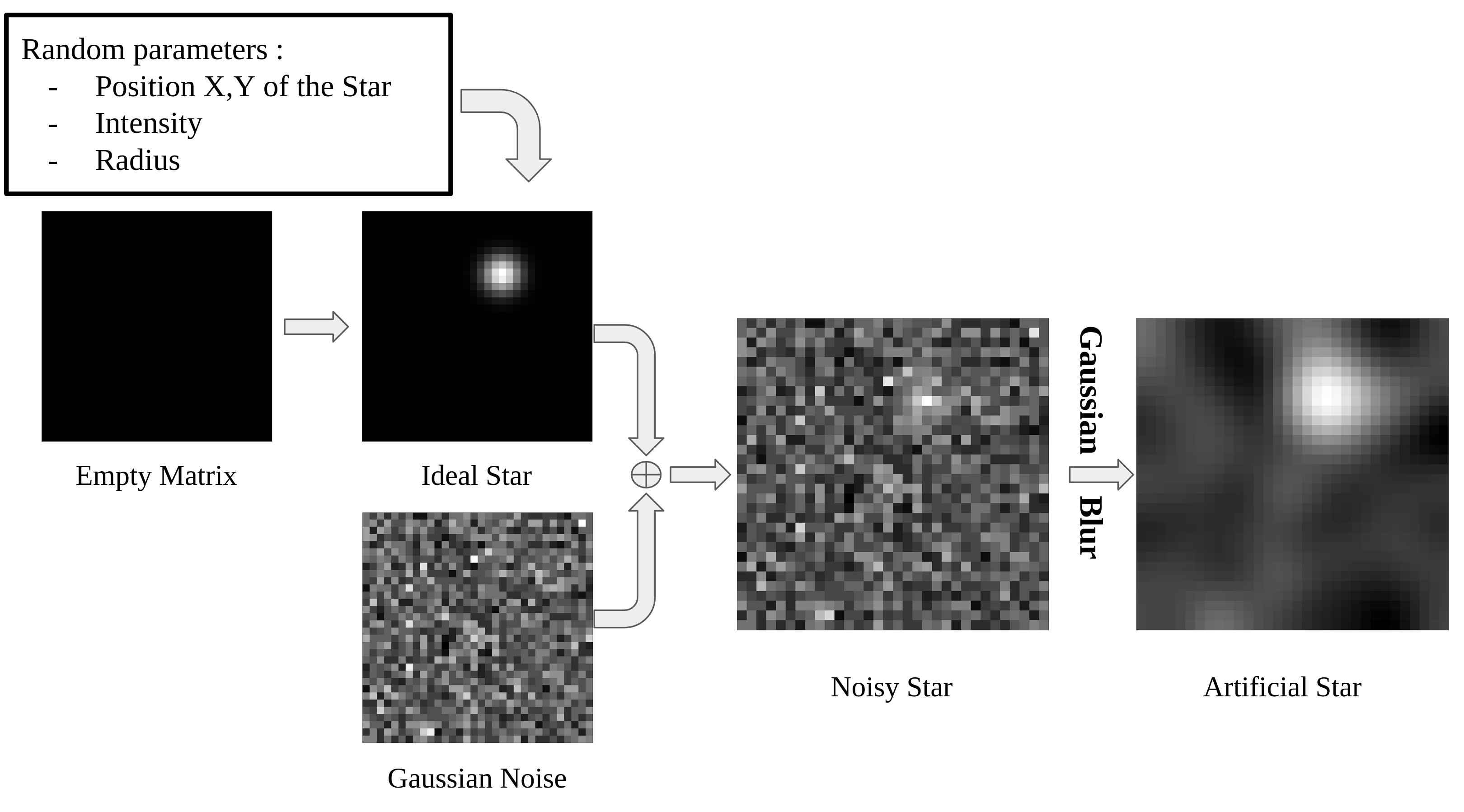}
   \includegraphics[width=4cm]{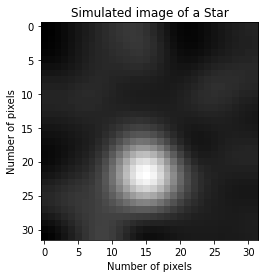}
   \includegraphics[width=4cm]{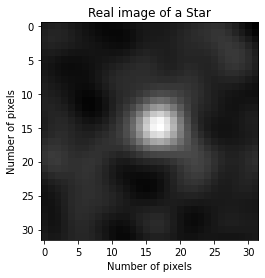}
      \caption{Algorithm to create a random artificial star snippet as part of the training set (left) and comparison with TYC 6812-00410-1 as seen with an eVscope (right). Display is at the same scale for both frame for comparisons.}
         \label{FigAlgoStar}
         
\end{figure*}

To construct a star as seen by an eVscope that has been convolved with a Gaussian filter (as in Figure \ref{FigAlgoStar}), we begin by defining that a star has a background of zeros with a zero matrix $\mathcal{O}_{32 \times 32}$. 
We chose this size of snippets because it readily fits an average star's PSF even if the star is not perfectly centered. 
For the noise distribution we tested and compared different models with Unistellar's database, showing that the best model of light noise is the Poisson noise, which correspond to the noise due to a low light intensity. We decide to neglect the Gaussian noise created by the electronic readout as its order of magnitude is far lesser than the Poisson noise \citep{lanteri2005restoration}. The star's span varies too, determined by the variance ($\sigma$) value of the Gaussian blur, from $\sigma=1$ to $\sigma=4$, so the more we blur, the more the span increases.

We obtain the following results for the simulation of a star, as seen in the Figure \ref{FigAlgoStar}. Figure 3 shows that the artificial star is morphologically similar to the observed point spread function, validating our algorithm to create an appropriate training set.

The next step is to create a large data set of observations, labeled ``true'' or ``false'', corresponding respectively to ``this set contains an occultation'' and ``this set does not contain an occultation''.
The simplest way to create an artificial occultation on a set of [N] snippets-- that we call a sequence--is to create a function that generates two stars with random noise. See Appendix \ref{appendix:Terminology} for more information on the used terminology.
For a positive observation (i.e., a visible occultation), the targeted star (TaSt), disappears during a random number of frames while the reference star (RfSt) does not.
For a negative observation (i.e., no occultation), the two stars remain clearly visible on top of the background. Note that their magnitude can vary slightly during the sample of frames but will never fade.
Also, we created the stacks of TaSt and RfSt such that the background level has the same intensity, since the two stars are supposed to be in the same frame within several arcminutes of each other.
A first model is trained and tested on these two cases. However, real-life data tends to include more complex photometric variation than a star simply maintaining a constant brightness or disappearing entirely, and this complexity is poorly described with these two extremely simple samples. 
For example, a cloud passing across the FoV would create a variation of star magnitude that is not an occultation, thereby creating a false result.
To solve this problem, we enriched the training cases by adding the following patterns to the training database, according to our experience with different kinds of data generated by eVscopes. The cases herein are considered as positive observations: 
\begin{itemize}
    \item The TaSt fades before disappearing. This can occur with large or slow asteroids: they gradually decrease the magnitude of the star on several frames before the star reaches its lowest magnitude
    \item The TaSt does not disappear entirely (its brightness drops from 100\% to $\sim$10\%). This happens with brighter asteroids, or when the asteroid’s magnitude is comparable to the star’s magnitude. We decided to use 10\% as the occultation floor as it is enough to consider it is due to an asteroid and not too high so fluctuation in the star's signal would not be considered an occultation.
    \item The TaSt disappears but does not reappear (long occultation). Because we are feeding ODNet with a finite number of frames, a long-duration occultation would not be seen as an occultation if we did not teach the CNN that it is an occultation.
    \item The TaSt is absent at the beginning of the set but appears at the end of the occultation. This is the similar case as in the previous item but for the reappearance.
\end{itemize}
The following should be considered to be negative observations:
\begin{itemize}
    \item The TaSt and the RfSt both disappear at the same time. This can be caused by a cloud or by wind shacking the telescope, affecting both stars at the same time.
    \item The RfSt is “occulted” (a disappearance following a pattern of the TaSt). The RfSt is not suppose to be occulted, but if a cloud covers this star before the TaSt, the CNN must understand that it is not an occultation.
    \item No stars on the FoV.
    \item Both stars fade gradually before disappearing. Once again, this can be caused by clouds.  
    \item The RfSt, the TaSt, or both disappear sporadically on frames of the sample. This is typical of a very windy observation, or if the stars fall out of the FoV or if some frames are missing. Also, this can happen with variable background noise, as when the citizen scientist is on the side of a road and car lights randomly pollute the FoV. See the Appendix \ref{appendix:Artificial} for examples.
\end{itemize}

We are now able to generate a data set of [K] observations, being composed of two sets of [N] snippets--32 $\times$ 32 pixels--for both TaSt and RfSt.

\begin{figure*}[!ht]
   \centering
   \includegraphics[width=\textwidth]{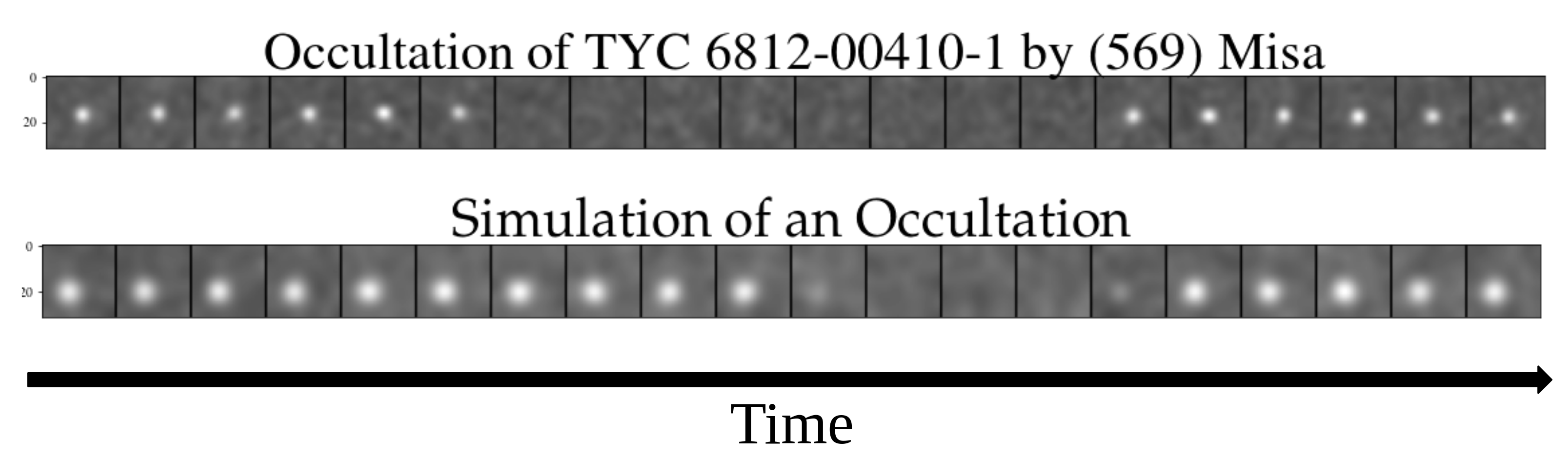}
      \caption{A simulated stack (bottom) made of 20 snippets including a occultation lasting 5 frames. A stack (top) based on a recorded occultation observed on May 27 2021 with the disappearance of the star over 8 frames. Those stacks are part of the sequences input to ODNet.
              }
         \label{FigCompareSet}
         
\end{figure*}

\subsection{Proposed Method} 
We propose a CNN to detect asteroids occultations from blurred raw frames.

\begin{figure*}[!ht]
   \centering
   \includegraphics[width=\textwidth]{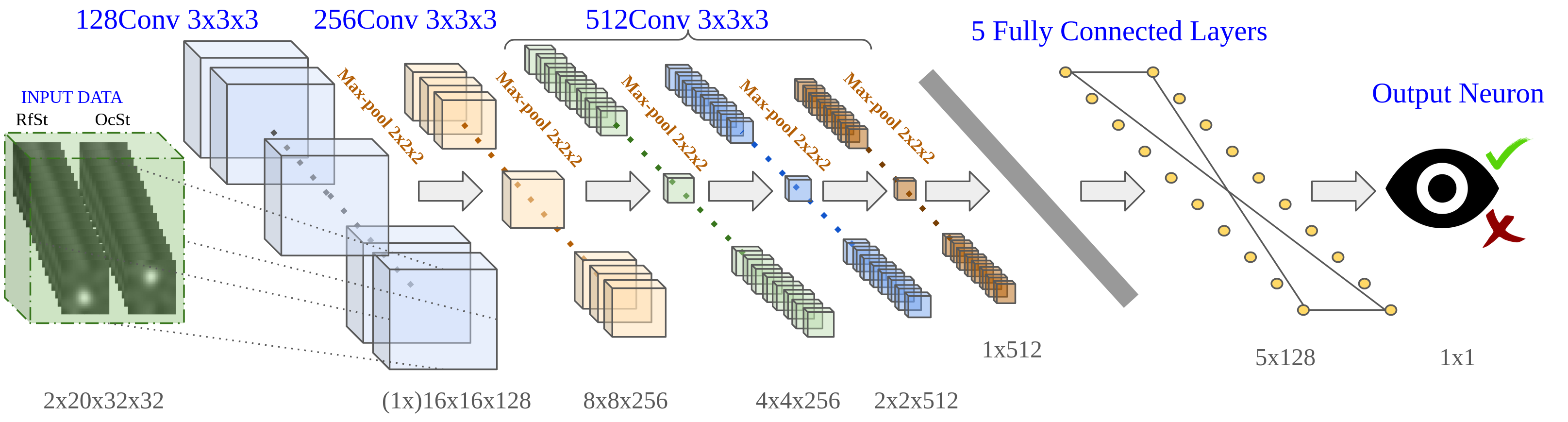}
      \caption{Visualization of ODNet's model chosen. We can observe the 4D data ( $2 \times 20 \times 32 \times 32 $) being flattened by the consecutive convolution and max-pooling layers, then, the 5 layers of neurons treat the sample to conclude if it contains or not an occultation. The user is informed of the output via one last neuron. The output neuron is a number from 0 to 1, seen as a probability of the event ``presence of an occultation in the set presented''. Between each layers of convolution, there is a 3D max-pooling, of size 2x2x2.}
         \label{FigModelSch}

\end{figure*}

Occultations are temporally limited events, necessarily composed of a single star's disappearance, a time when it is obscured and its reappearance. Therefore it is not possible to detect such event precisely from a single frame. Also, a drop of flux intensity may sometimes be caused by various external parameters (e.g. sensor gain, presence of clouds), making real occultation detection even more challenging.\\

In order to overcome the brightness variation, the observed star is at first compared to a second one, a reference star, close to the field of view and bright enough, to ensure a flux ratio as independent as possible from external phenomena. 
Second, a set of N=20 snippets rather than a single frame is more appropriate to take into account the temporal evolution of magnitude of both stars. A rolling sample of twenty frames is fed to the neural network, permitting the CNN to ``scan'' the entire observation for occultation. Note that the sample is normalized on the 20 stars, meaning that if the occultation appears, we will be able to see the star disappearing on the same scale, like on Figure \ref{FigCompareSet}. We cannot apply the CNN on the entire set of frames in an observation, because each observation has a different number of frames and our input must be standardized.\\

In that regard, inputs to the neural network should be two pre-processed stacks of stars (N snippets). The first stack contains the supposedly occulted star taken consecutively. The second stack contains snippets of a reference star extracted from the same frame taken consecutively. This second stack will help the network understand if the change in intensity in the first stack is an occultation. 
 Also, the convolutions will not be applied to a single frame but rather to the sequence of frames, meaning that we are looking for a pattern within a frame and between frames.
 The pattern sought within the snippet is the star, and the temporal pattern sought between frames is the disappearance of stars, if it happens. These patterns are illustrated in the Appendix \ref{appendix:Artificial}.
Afterward, we use a set of fully connected neuron layers to extract the final information out of the convolutional layers: the pattern information must be transformed into a “Yes/No” answer to the question “ Does this sequence contains an occultation?”

First, as in the legacy method, we apply a Gaussian blur of $2\sigma$ to each snippet in order to round the stars and diminish the background noise. Second, the citizen's blurred observation is run through an algorithm to determine where the star and the reference star are located, and to reduce each frame to a set of 2 snippets, each 32 $\times$ 32 pixels. Thus, on this model, we are able to treat the occultation of each star in turn: this will be further described in Section \ref{Discussion}.
\newline
In our model, as described in Figure (\ref{FigModelSch}), we use 5 layers of 128 neurons using a Rectified Linear Unit (ReLU, equation \ref{Relu}), and the output neuron is with a sigmoid activation function (equation \ref{sigmoid}), in order to emit a probability of presence of the occultation as a number between 0 and 1. Before these dense layers, we are using 5 layers of 3D convolutions, coupled with max-pooling, permitting to reduce the size of the input sequence before they go in the dense layers. It produces a model with $\sim$18 millions parameters to train. To that end, we need to create a large training set that will fix the input for the desired output of ODNet. 

\begin{equation}
\label{Relu}
    ReLU(x) = max(0,x)
\end{equation}

\begin{equation}
\label{sigmoid}
    \sigma(x) = \frac{\mathrm{1} }{\mathrm{1} + e^{-x} }
\end{equation}

\subsection{Experiments}
To train our model, we used a Stochastic Gradient Descent  with a of learning rate=$0.001, \beta_1=0.9, \beta_2=0.999, \epsilon=1e-07$, and the Adam Optimizer \citep{kingma2014adam}.

The training set is composed of 80\% of the simulated data, and the test set of the remaining 20\%. After the training, we observed that a good trade-off between [K] and the efficiency of the network was to use K = 50\,000 observations, half negative, half positive, following the different photometric cases explained in the Table (see annexes \ref{appendix:Artificial}). As a reminder, it represents 2\,000\,000 snippets of stars randomly generated and variable in position in the frame, span, and S/N.

\begin{figure}[!h]
   \centering
   \includegraphics[width=7cm]{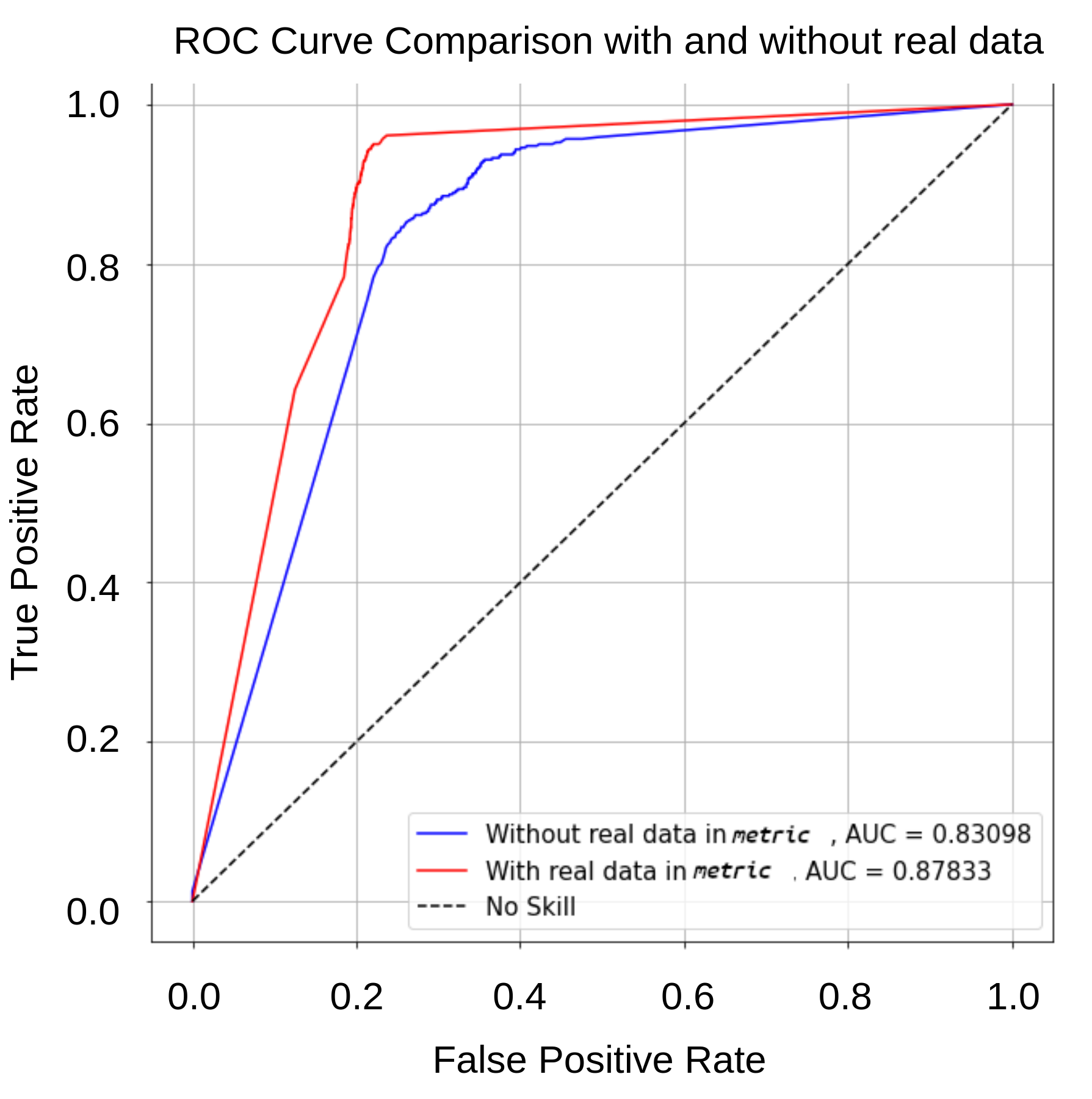}
            \caption{ROC curve of model including real data during the training and including only artificial data. A perfect model would be represented as a right angle at the top left corner: we see that as we input only five real observations in the test set, our AUC (area under curve) goes from 83\% to 87\%. Thus, more real data permits a more precise CNN. Note that this ROC curve is not from the same model as the model in the one in Table \ref{table:metric}, it is an example for the enhancement of the model depending on the training data.
              }
         \label{FigROC}
   \end{figure}

With only simulated data, the efficiency of our CNN was unsatisfactory, because the test-set was generated in the same way as the training set. To remedy this situation, we added some real observations during the testing, using them only as a metric: five positive occultations have been input at the end of each optimization's iteration, or ``epoch'', which tells us if the model is improving over time. This way, after each epoch, we would not determine our precision on the simulated data anymore. This test set represents $\sim$6\,000 frames, with $\sim$200 frames containing the occultation event. The positive data representing only 3.33\% of the test set, we need to balance the way the metric is calculated. We gave thirty times more value to the frames containing the occultation so the total of positive frames has the same weight as the negative frames. Better results were immediately observed, as it can be seen on the ROC \citep{hoo2017roc} curve in Figure \ref{FigROC}.

\section{Test on Real Data}

For the evaluation of our model, we selected sixty-six observations: twenty-four positives and forty-two negatives, representing different S/N, star shapes and occultations duration. It represents $\sim$30,000 sets of pair of stars, among them $\sim$450 sets containing a positive observation with the rest being negative.
The average execution time of ODNet is one second per hundred frames. 

Let us take the example of (617) Patroclus observed by a citizen astronomer on September 5, 2021. This observation took several hours for humans to investigate because it needed background reduction and a manual selection the area we wanted to study: clouds and wind made the observation seems--seen by the legacy code--to contain three occultations before the real one. ODNet needed only three minutes: three minutes to extract the two sets of stars' snippets, and less than eight seconds to give us its output. \\

\begin{figure}[!b]
   \centering
   \includegraphics[width=17cm]{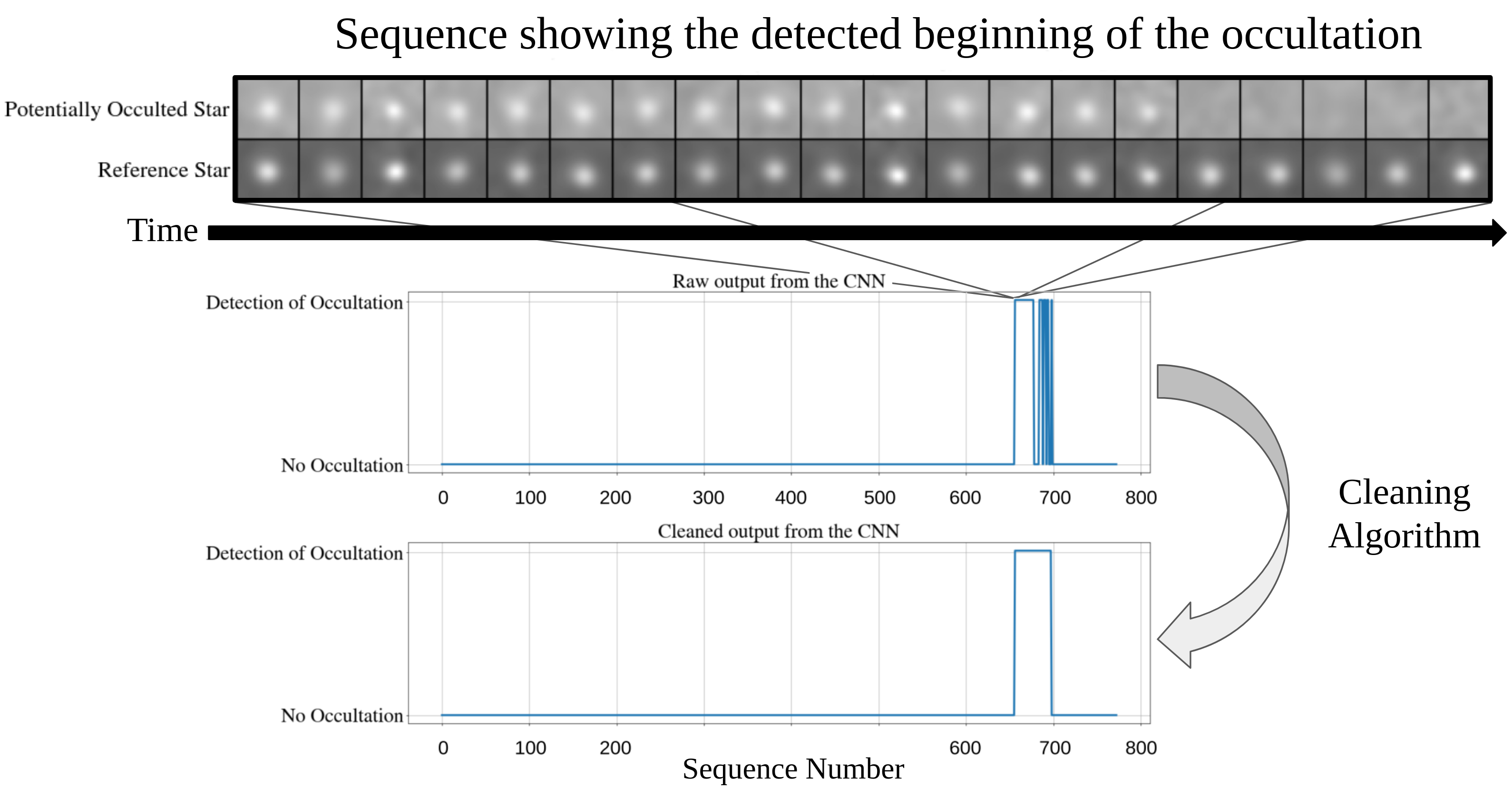}
      \caption{Patroclus's observation seen by ODNet. From top to bottom: star occulted - reference star - Raw output from ODNet - Cleaner Output.}
         \label{FigPatro}
\end{figure} 

As we can see on Figure \ref{FigPatro} (middle panel), occasionally during the occultation (i.e., where the detection plateaus), some samples are detected negative. We coded a short algorithm that interpret singular points as natural errors: a single positive point with no other positive around it is declared negative, and a negative surrounded by positives is changed by to positive.
These singular outliers should not exist in most standard single-body occultations given the scanning technique we use to search the data set. Even if an occultation occurs in just a single frame, ODNet will detect that occultation each time it appears in the set of 20 adjacent frames (i.e., 20 times). Very specific cases of binary asteroids or ring systems could possibly be hidden by this technique, identified as just a standard single occultation. Identifying these cases are not the initial purpose of ODNet, so we are not concerned by this presently (see Section~\ref{Discussion}).  

To rate our model, we used five metrics: the precision, the recall, the F1, the accuracy and the time of execution. They are calculated as follows:\newline

\begin{table}
\caption{Metric Comparison of the Different Methods}
\label{table:metric} 
\begin{tabular}{cc}
\begin{minipage}{0.5\textwidth}
    \begin{flushleft}

        \begin{tabular}{c c c}

        \hline\hline
         Metrics & Legacy Code & ODNet \\
        \hline
         Precision & 64.7\% & \textbf{91.3\%} \\
         Recall & \textbf{91.7\%} & 87.5\% \\
         F1 & 75.9\% & \textbf{89.4\%} \\
         Accuracy & 78.8\% & \textbf{92.4\%}\\
         Total time & 20 minutes & \textbf{2-4 minutes} \\
        \hline
        \end{tabular}
        
  \end{flushleft}
\end{minipage}&
\begin{minipage}{0.5\textwidth}
    \begin{equation*}
        \begin{aligned}
        \\
            Precision &= \frac{True Positive}{True Positive + False Positive} \\
            Recall &= \frac{True Positive}{True Positive + False Negative} \\
            F1 &= 2 \times\frac{Precision \times Recall}{Precision + Recall} \\
            Accuracy &= \frac{True Negative + True Positive}{Total} \\
        \end{aligned}
    \end{equation*}
\end{minipage}

\end{tabular}
\end{table}

Precision describes the ability of the CNN to correctly detect occultations without errors of positive label on a given set.
Recall describes the ability of the CNN to avoid mislabelling a true occultation as negative.
The F1 Score represents the balance of the CNN to as much correctly detect the occultation, but also not to miss or miss-label them.
Accuracy permits us to score the ability of the CNN to generate correct detections (true negatives or true positives) out of the entire test set.
With ODNet, we achieve a precision of 91.3\% and a recall of 87.5\%, a F1-score of 89.4\% with an accuracy of 92.4\%. These values exclude the ``inconclusive'' case, where the observation is not exploitable due to environmental factors.

We can see in Table \ref{table:metric} that the deep-learning method competes successfully with the legacy code in terms of precision, F1, and accuracy, but is less effective in terms of recall.
This can be understood as the CNN making far fewer errors when it comes to labeling negative observations as negative: it generates fewer false positive. Nevertheless, it has missed some occultations (positive detections), while the legacy method did not. The reasons for these missed detections are known and will be considered in the Section \ref{Discussion}.

Also, the CNN is more efficient in terms of time: a typical extraction of data to make the sample of stars is 82 seconds per 1000 frames and the CNN itself takes less than a minute. As example, the legacy method takes twenty minutes to generate a conclusion for the same kind of occultation. We have, in short, developed a faster and more efficient method. It is important to note that 1000 frames represent generally five minutes of observational time, so as long as ODNet's time of response is faster than the time it takes for the data to be observed, we can imagine real-time detection. 

\begin{figure}
   \centering
   \includegraphics[width=15cm]{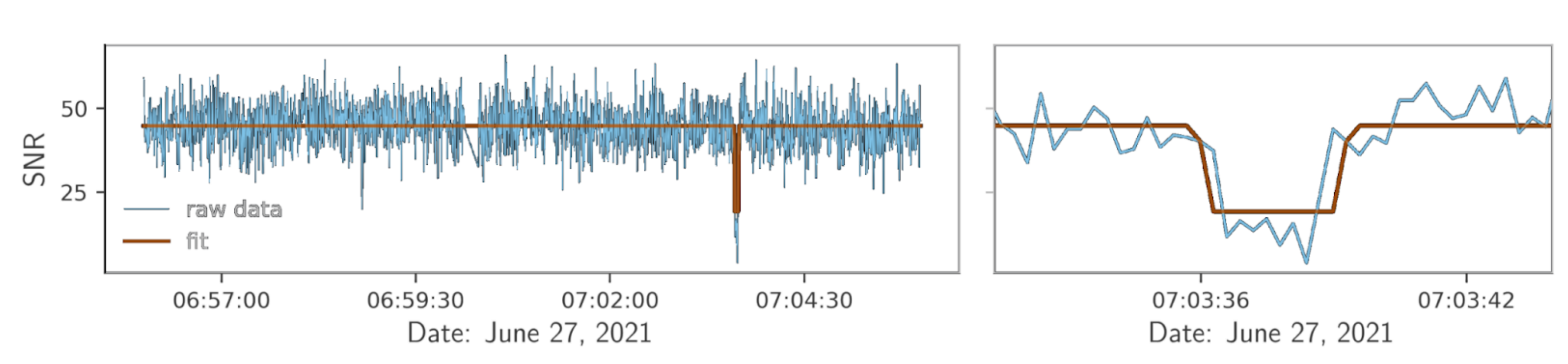}
   \includegraphics[width=12cm]{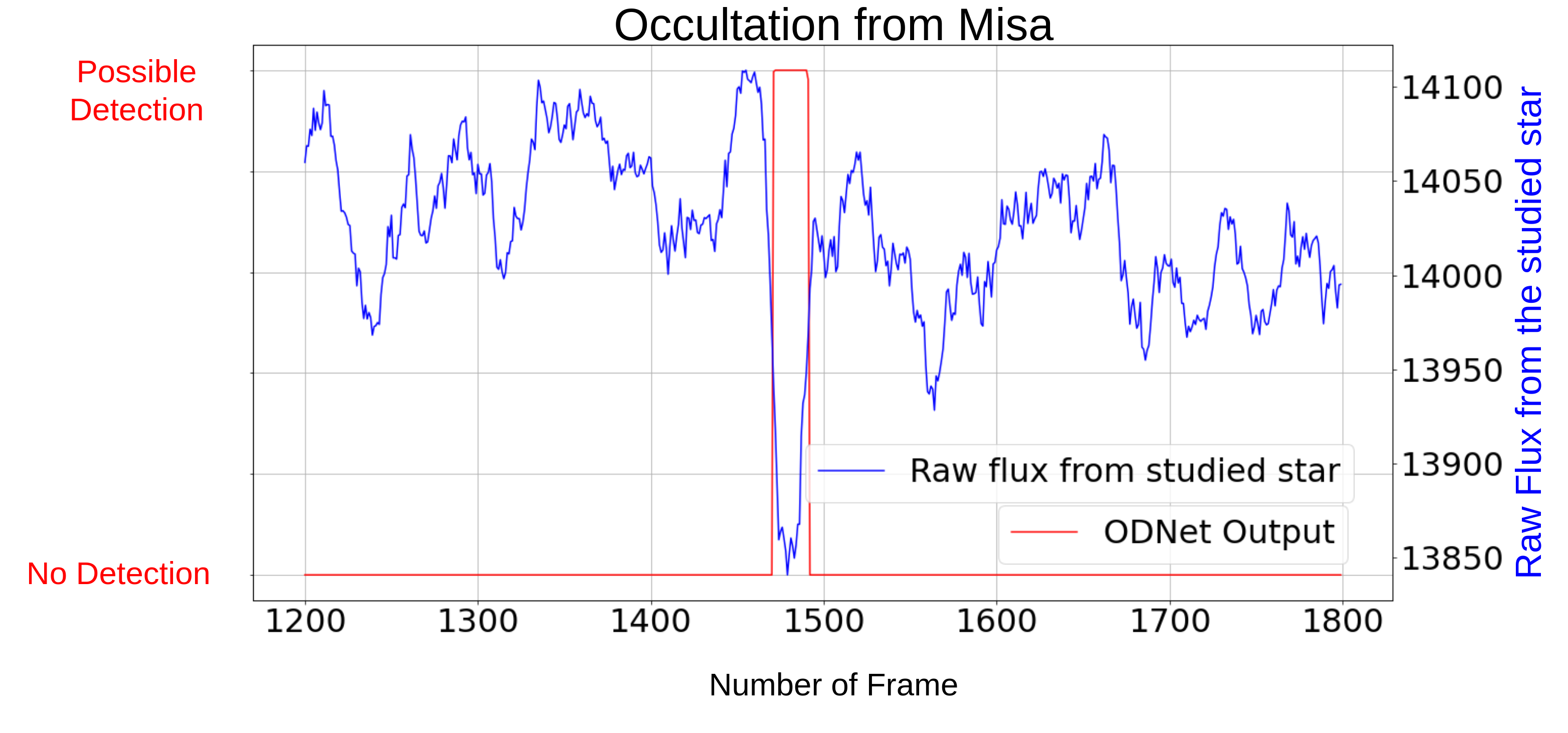}
      \caption{Public occultation report for Misa seen by the legacy code [top] and the same observation seen by ODNet [bottom]. We can see that on simple occultation (no wind nor cloud), ODNet performs as good as the legacy code
              }
         \label{FigMisaML}
\end{figure}

Another example of detection of an occultation is the observation of Misa we used in Figure \ref{FigPhotoMisa} for the description of the legacy code. We can see on Figure \ref{FigMisaML} that the algorithm is indeed detecting the occultation in the same spot as the legacy method.\newline

\section{Discussion}\label{Discussion}

On the real observational data set that we used as a test, the magnitude of the studied star in G-band was from 8 to 13, with an average value of 12. We compare the results of ODNet and the legacy code within the limitations presented by the legacy code. Hence, we were not able to evaluate ODNet with observations presenting a S/N lesser than 20.
For stars fainter than $G=14$, the legacy code is not efficient anymore because the star signal is lost into the background.
For example, in occultation of UCAC4 669-035752 shown in Appendix \ref{appendix:BadStar}by the main-belt asteroid (795) Fini on March 10, 2021, a citizen astronomer was observing from downtown of a large city in Japan. Due to the bright sky, the S/N of the observed star was about 20 during the whole observation.
Neither the legacy code nor the Neural Network algorithm could reach a conclusion using this observation. In this case, the cleaning algorithm shown in Figure \ref{FigPatro} will be unable to find the signal of unique occultation, it will consequently label this observation as ``Inconclusive'' since it has not been able to conclude on the presence or absence of an occultation. \newline

We noted that most of the false results were caused by to background noise, clouds, or very low S/N. ODNet is faster and better at detection than our legacy code (as seen in Section \ref{section:2}), but problems created by environmental conditions such as poor weather or light pollution cannot yet be solved.
In the future, we could create a better training set which will include simulation of cloud perturbations so the model is better prepared to identify such a case and label it as ``inconclusive''.

ODNet is fast but has some limitations.
First, a bright asteroid's occultations cannot be detected with the current deep-learning method because its apparent magnitude is too close to the magnitude of the occulted star. We can take the example of the observation involving the main belt asteroid (13) Egeria on the night of March 15-16, 2021.
The asteroid has occulted UCAC4 655-038057, a magnitude $G = 12.5$ star, while Egeria has a predicted magnitude 11.6 in V-band, so the occultation's drop was predicted to be less than 1. In this case, ODNet is unable to detect the event. An algorithm similar to the legacy code is able to detect such a shallow occultation.\newline

We could enhance and improve the capability of the CNN algorithm by making it more versatile and aware of the environment. For instance we could make the output of the network multi-class, so adding more than one probability. Transforming one ending-neuron to a list of labelled neurons, so we can collect more details on the quality of the recorded frames. For instance, we could add a ``cloud'' neuron, which would determine the probability that the frames are contaminated by clouds, or a ``low S/N'' neuron, or even a detector of vibrations and wind. This improved ODNet will return a metric on the quality of the data, helping to assess if an occultation is detected along the observation.  \newline

The method described in this article is not costly in term of computing power, opening the possibility for us to embed it on the telescope directly. Indeed the eVscope is doing the AFD on the first frame, and tracking the motion of the stars, meaning that once it is set, we can assume that the relative position of the studied star on the FoV is almost constant. A short code can be written to track a given star on each frames knowing its initial position, making us save precious time of calculation instead of doing AFD on each frames to keep track of TaSt and RfSt position. Also, due to the really simple operations that constitute the neural network, the integration of such computing method should be doable onboard a Unistellar telescope. \newline

Using the Unistellar's citizen scientist network, we are now able to process a large amount of data.
A typical preparation of data and ODNet's analysis is made within three minutes. By way of contrast, the legacy code takes 20 minutes to generate a conclusion for the same occultation. Clearly, we have achieved a faster and more efficient method. We believe we have created the first method of occultation detection using a Convolutional Neural Network. The results with this first algorithm are very encouraging and at low computational cost. Combined with the availability of citizen scientists, this CNN algorithm is well fitted to handle the large amount of data continuously generated by the Unistellar network. \newline

To date, ODNet is included in the scientific pipeline of the occultation analysis as a trust-worthy indicator for occultation detection: it allows us to focus our time and computing power on observations for which positive occultation events are indeed present and to know immediately, without deep diving into the observational data, if a report from the legacy code is correct or not. The legacy code is still needed to derive an accurate timing for disappearance and reappearance with their uncertainties.
The next step could be to embed this ML model directly into the telescope, doing the calculation during the acquisition of data (edge-computing) and flagging the data before it is even sent to the SETI astronomers. We believe that ODNet can be used with other telescope data as long as the sequence of two stack of twenty snippets (32 by 32 pixels) is provided to the code as an input. The code and few examples are available on \href{https://github.com/doriancazeneuve/ODNet}{GitHub\footnote{For printed version : https://github.com/doriancazeneuve/ODNet}}.

\section{Perspective}

This study was applied to known events in that we knew which star to look at and which star to use as a reference. But what if we take a random observation on the database without knowing which star to observe? 
Due to the short duration of this study, we can simply select tens of stars inside a frame, declare one as the reference star and the others as potentially occulted, and let ODNet do the rest.
But another enhancement could use more recent CNN techniques to detect occultations: fast-RCNN (\cite{girshick2015fast}) and YOLO (\cite{redmon2016you}) models could be applied to the entire frame, detecting what is a star and, among them, which ones are occulted. It would be a more of an end-to-end application as the direct output of the observation will be an input to the AI, and the last output would be the beginning and end of the occultation (if there is one) without any human pre- or post-treatment. The direct application of such enhancement could be the detection of unpredicted asteroid occultations. For example, this could open the field to autonomous detection of unknown, or poorly known, NEAs, or even distant Transneptunian objects.
\newline

\section{Acknowledgements}
\begin{acknowledgments}

This material is partially based upon work supported by the National Science Foundation under Grant No. 1743015, and the generous donation from the Richard Lounsbery foundation and the Moore Foundation. \\
P. Dalba acknowledges support from the 51 Pegasi b Fellowship funded by the Heising-Simons Foundation.

\end{acknowledgments}

\software{ \texttt{Tensorflow}, \citep{tensorflow2015-whitepaper} \\
\texttt{Photutils}, \citep{10.5555/1593511}}

\newpage


\appendix

\section{Example of low S/R star seen from an eVscope}
\label{appendix:BadStar}

\begin{figure}[!h]
   \centering
   \includegraphics[width=6cm]{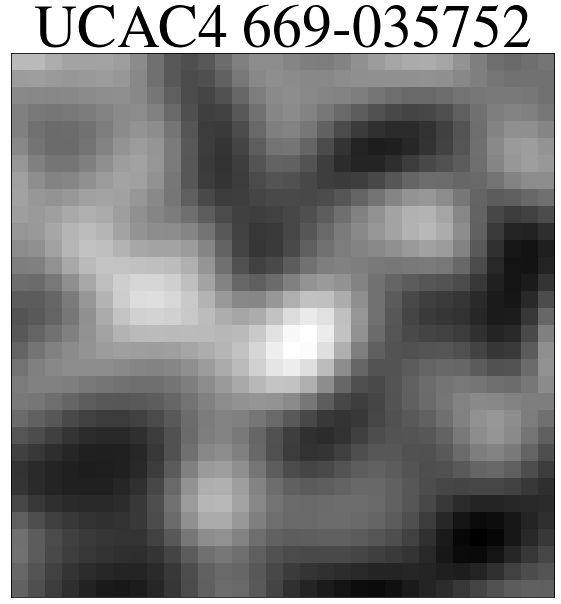}
      \caption{Example of low S/N star, mixed with its background. The star is at the exact center of the frame.
              }
\end{figure}

\section{Terminology for simulating ODNet's input}
\label{appendix:Terminology}

\begin{figure}[!h]
   \centering
   \includegraphics[width=15cm]{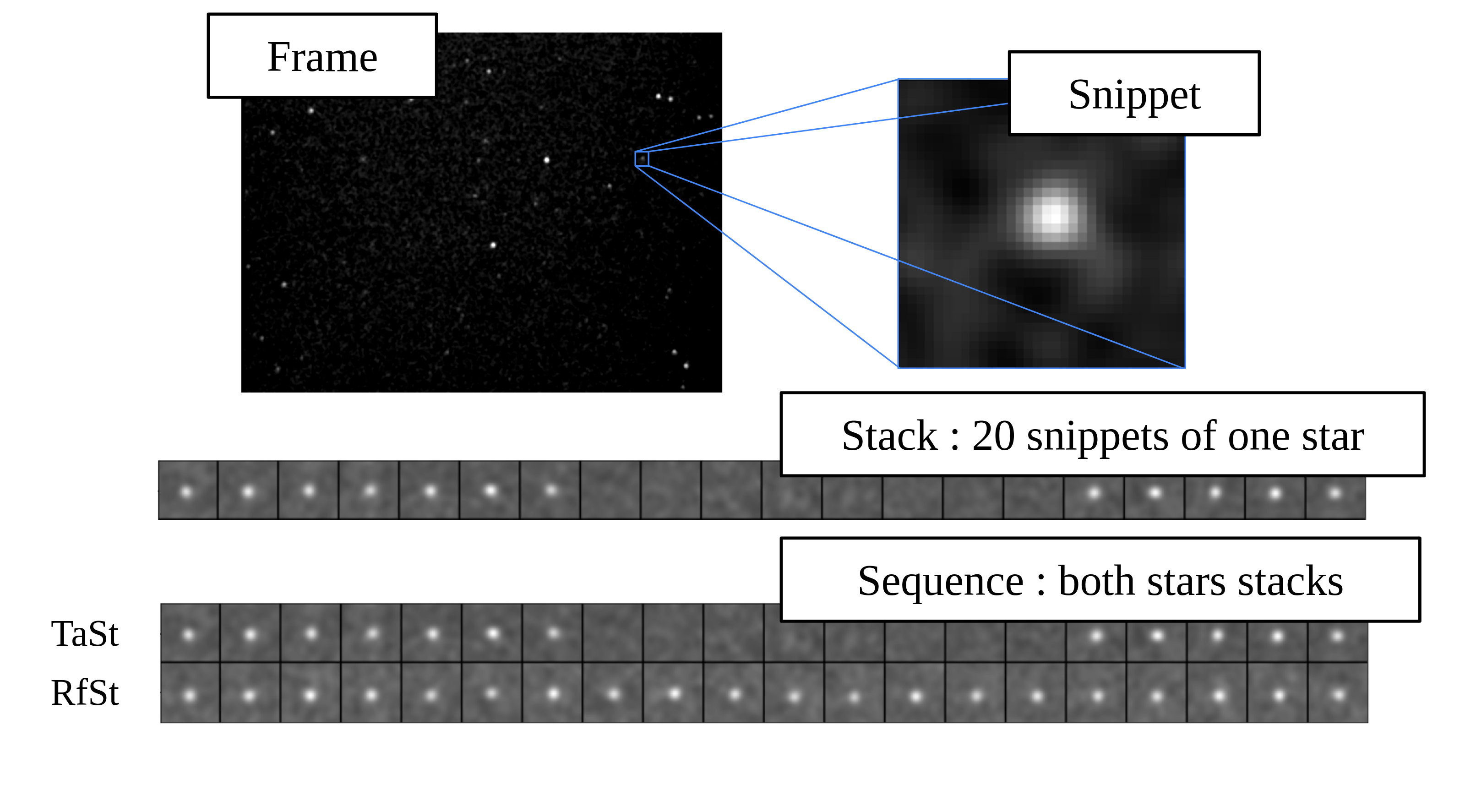}
      \caption{Terminology of the simulated set. A frame contains the stars in the FoV of the eVscope. The snippet is an image containing a star, artificial or not. A stack is a set of N snippets, permitting to look at the evolution of a star through time. A sequence is a combination of the TaSt and RfSt stacks: this is the input that is fed to ODNet.
              }

\end{figure}

\section{Zoology of photometric signals used for model's training }
\label{appendix:Artificial}

\startlongtable
\begin{deluxetable}{llll}
\tabletypesize{\scriptsize}
\startdata   
\tablehead{
\colhead{Name} & \colhead{Stars} & \colhead{Description} & \colhead{Visual Representation}} \\
\multirow{2}{2em}{P1} & TaSt & Disappear abruptly &  \raisebox{-\totalheight}{\centering}{\includegraphics[width=9cm]{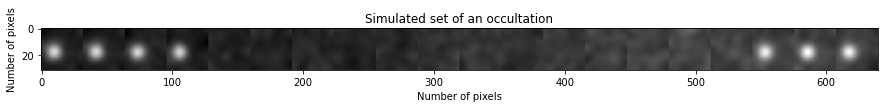}} \\ 
& RfSt & Stay constant & 
\raisebox{-\totalheight}{\centering}{\includegraphics[width=9cm]{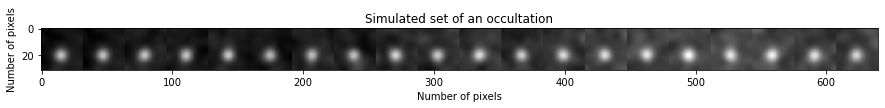}} \\ 
\hline
\multirow{2}{4em}{P2} & TaSt & Fade on one frame before disappearing & \raisebox{-\totalheight}{\centering}{\includegraphics[width=9cm]{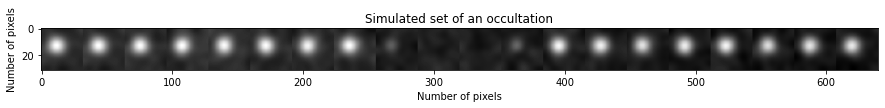}} \\ 
& RfSt & Stay constant & \raisebox{-\totalheight}{\centering}{\includegraphics[width=9cm]{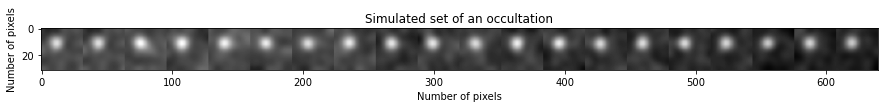}}\\ 
\hline
\multirow{2}{2em}{P3} & TaSt & Disappear abruptly and does not reappear (long occultation) & \raisebox{-\totalheight}{\centering}{\includegraphics[width=9cm]{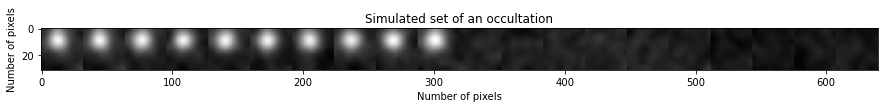}} \\ 
& RfSt & Stay constant  & \raisebox{-\totalheight}{\centering}{\includegraphics[width=9cm]{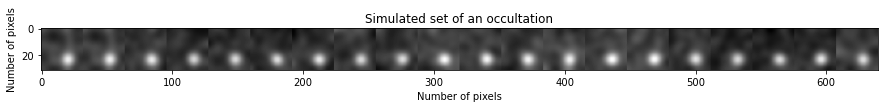}} \\ 
\hline
\multirow{2}{4em}{P4} & TaSt & Appear abruptly & \raisebox{-\totalheight}{\centering}{\includegraphics[width=9cm]{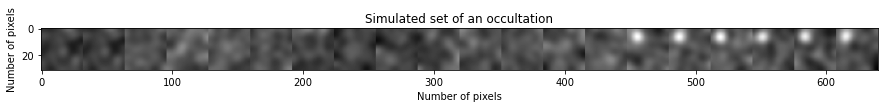}} \\ 
& RfSt & Stay constant & \raisebox{-\totalheight}{\centering}{\includegraphics[width=9cm]{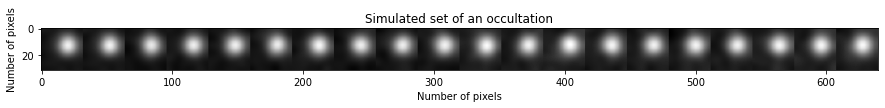}} \\ 
\hline
\multirow{2}{4em}{P5} & TaSt & Losses 90\% of its magnitude & \raisebox{-\totalheight}{\centering}{\includegraphics[width=9cm]{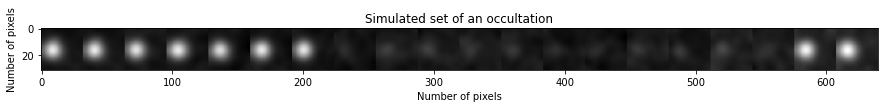}} \\ 
& RfSt & Stay constant & \raisebox{-\totalheight}{\centering}{\includegraphics[width=9cm]{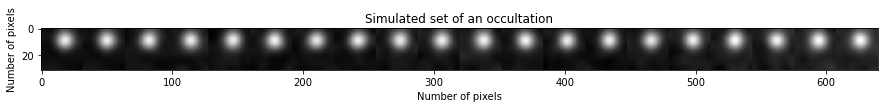}} \\ 
\hline
\multirow{2}{4em}{P6} & TaSt & long occultation, no star & \raisebox{-\totalheight}{\centering}{\includegraphics[width=9cm]{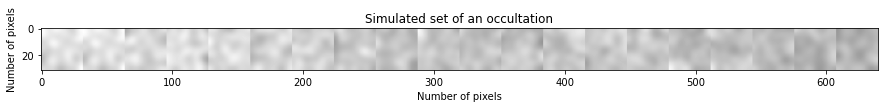}} \\ 
& RfSt & Stay constant & \raisebox{-\totalheight}{\centering}{\includegraphics[width=9cm]{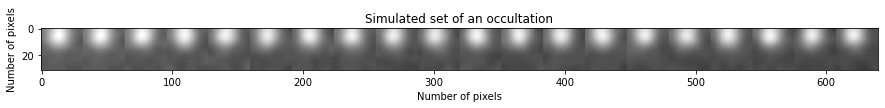}} \\ 
\hline
\multirow{2}{4em}{N1} & TaSt & Stay constant & \raisebox{-\totalheight}{\centering}{\includegraphics[width=9cm]{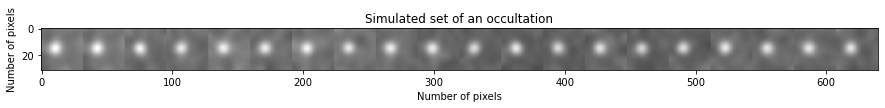}} \\ 
& RfSt & Stay constant & \raisebox{-\totalheight}{\centering}{\includegraphics[width=9cm]{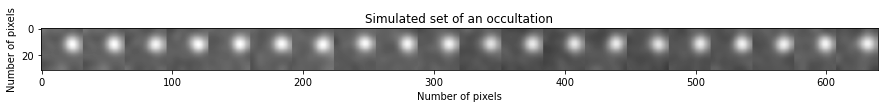}} \\ 
\hline
\multirow{2}{4em}{N2} & TaSt & Disappear one one frame & \raisebox{-\totalheight}{\centering}{\includegraphics[width=9cm]{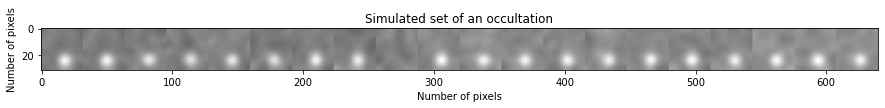}} \\ 
& RfSt & Stay constant & \raisebox{-\totalheight}{\centering}{\includegraphics[width=9cm]{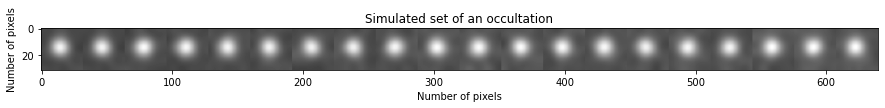}} \\ 
\hline
\multirow{2}{4em}{N3} & TaSt & Disappear at the same time (cloud) & \raisebox{-\totalheight}{\centering}{\includegraphics[width=9cm]{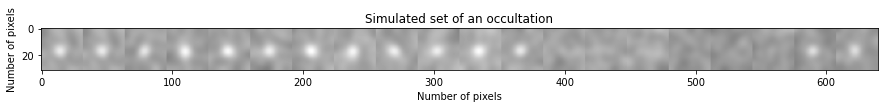}} \\ 
 & RfSt & Disappear at the same time & \raisebox{-\totalheight}{\centering}{\includegraphics[width=9cm]{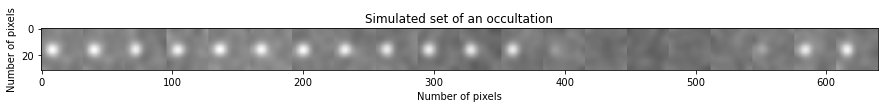}} \\ 
\hline
\multirow{2}{4em}{N4} & TaSt & Seems occulted (P5: High altitude cloud ) & \raisebox{-\totalheight}{\centering}{\includegraphics[width=9cm]{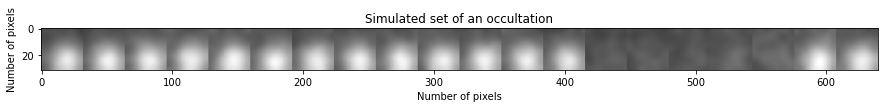}} \\ 
& RfSt & Seems occulted (P5: High altitude cloud ) & \raisebox{-\totalheight}{\centering}{\includegraphics[width=9cm]{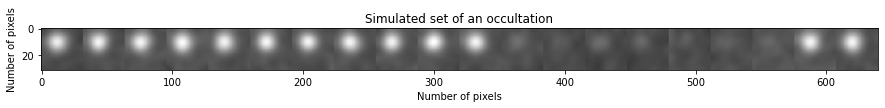}} \\ 
\hline
\multirow{2}{4em}{N5} & TaSt & No Star & \raisebox{-\totalheight}{\centering}{\includegraphics[width=9cm]{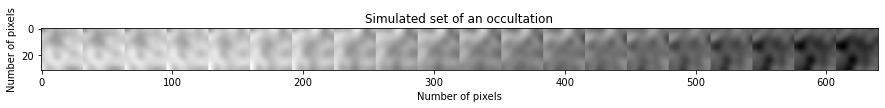}} \\ 
& RfSt & No Star & \raisebox{-\totalheight}{\centering}{\includegraphics[width=9cm]{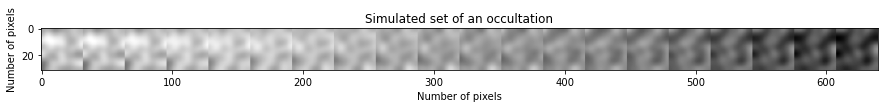}} \\ 
\hline
\multirow{2}{4em}{N6} & TaSt & Stay constant & \raisebox{-\totalheight}{\centering}{\includegraphics[width=9cm]{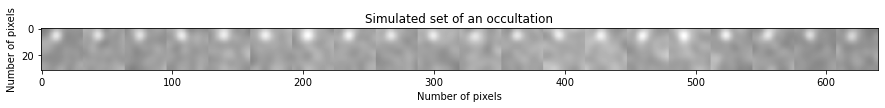}} \\ 
& RfSt & Blurred & \raisebox{-\totalheight}{\centering}{\includegraphics[width=9cm]{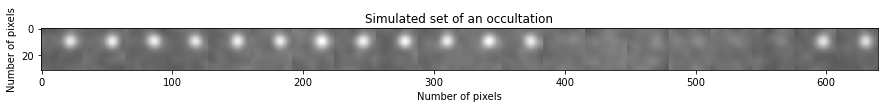}} \\ 
\hline
\multirow{2}{4em}{N7} & TaSt & Stay constant & \raisebox{-\totalheight}{\centering}{\includegraphics[width=9cm]{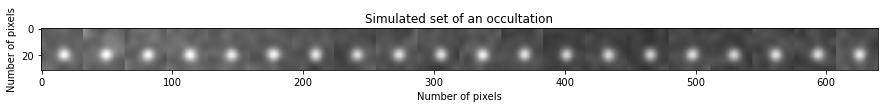}} \\ 
& RfSt & Occultation type P1 & \raisebox{-\totalheight}{\centering}{\includegraphics[width=9cm]{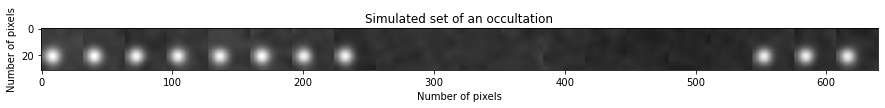}} \\ 
\hline
\multirow{2}{4em}{N8} & TaSt & Stay constant & \raisebox{-\totalheight}{\centering}{\includegraphics[width=9cm]{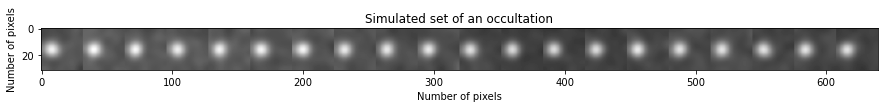}} \\ 
& RfSt & Occultation type P2 & \raisebox{-\totalheight}{\centering}{\includegraphics[width=9cm]{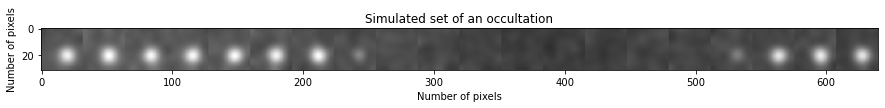}} \\ 
\hline
\multirow{2}{4em}{N9} & TaSt & Stay constant & \raisebox{-\totalheight}{\centering}{\includegraphics[width=9cm]{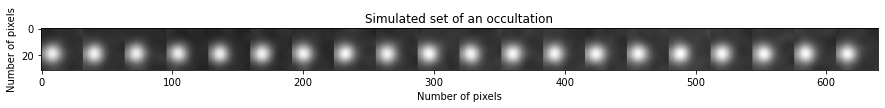}} \\ 
& RfSt & Occultation type P3 & \raisebox{-\totalheight}{\centering}{\includegraphics[width=9cm]{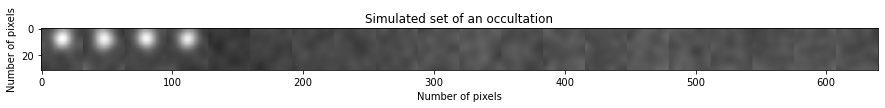}} \\ 
\hline
\multirow{2}{4em}{N10} & TaSt & Stay constant & \raisebox{-\totalheight}{\centering}{\includegraphics[width=9cm]{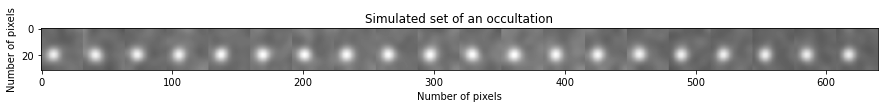}} \\ 
& RfSt & Occultation type P4 & \raisebox{-\totalheight}{\centering}{\includegraphics[width=9cm]{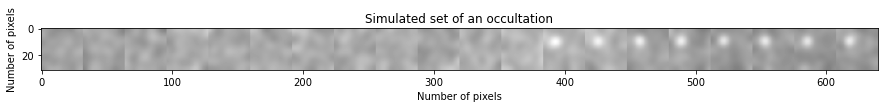}} \\ 
\hline
\multirow{2}{4em}{N11} & TaSt &Stay constant & \raisebox{-\totalheight}{\centering}{\includegraphics[width=9cm]{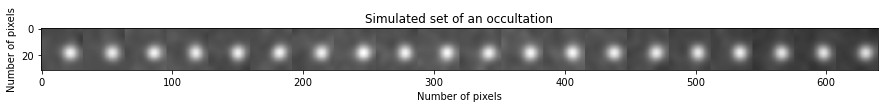}} \\ 
& RfSt & Occultation type P5 & \raisebox{-\totalheight}{\centering}{\includegraphics[width=9cm]{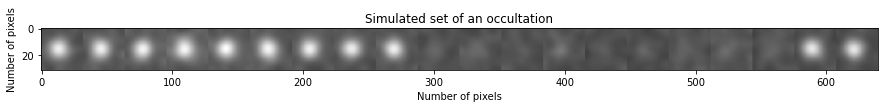}} \\ 
\hline
\multirow{2}{4em}{N12} & TaSt & Stay constant & \raisebox{-\totalheight}{\centering}{\includegraphics[width=9cm]{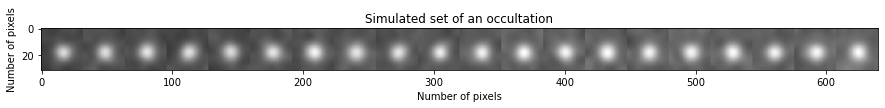}} \\ 
& RfSt & Occultation type P6 & \raisebox{-\totalheight}{\centering}{\includegraphics[width=9cm]{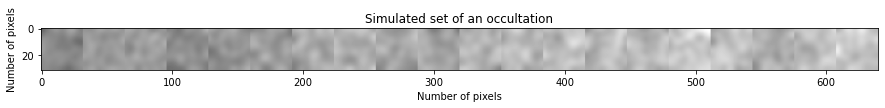}} \\ 
\hline
\multirow{2}{4em}{N13} & TaSt & Sporadic disappearance & \raisebox{-\totalheight}{\centering}{\includegraphics[width=9cm]{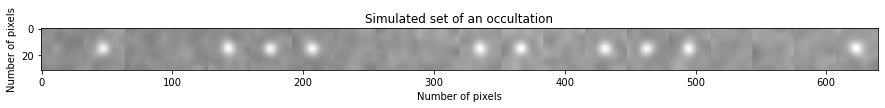}} \\ 
& RfSt & Stay Constant & \raisebox{-\totalheight}{\centering}{\includegraphics[width=9cm]{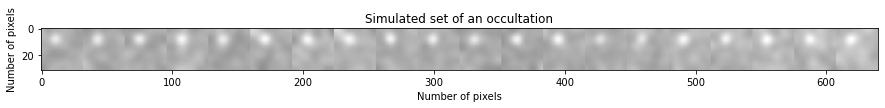}} \\ 
\hline
\multirow{2}{4em}{N14} & TaSt & Stay Constant & \raisebox{-\totalheight}{\centering}{\includegraphics[width=9cm]{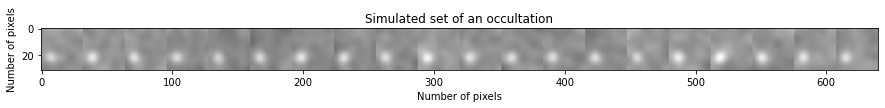}} \\ 
& RfSt & Sporadic disappearance & \raisebox{-\totalheight}{\centering}{\includegraphics[width=9cm]{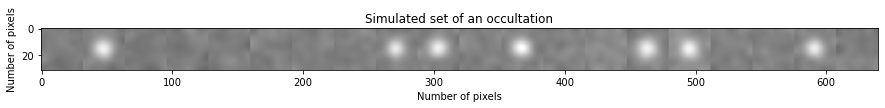}} \\ 
\hline
\multirow{2}{4em}{N15} & TaSt & Sporadic disappearance & \raisebox{-\totalheight}{\centering}{\includegraphics[width=9cm]{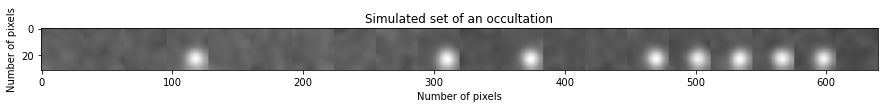}} \\ 
& RfSt & Sporadic disappearance & \raisebox{-\totalheight}{\centering}{\includegraphics[width=9cm]{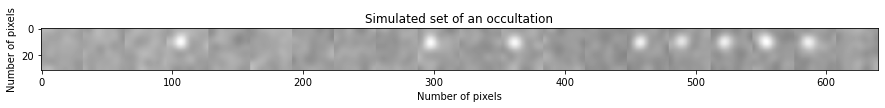}} \\ 
\hline
\enddata
\tablenotetext{}{Note that during the observation, the background intensity can vary, but the same way on the TaSt and RfSt, meaning that is the S/N of one star is affected, it is in the same way affected for the other. This way, environmental perturbations are not counted as anomalies but regular events that always happens. Such event is observed in real observations with high altitude clouds, or nearby punctual light pollution due to, as an example, car's light.}
\end{deluxetable}

\end{document}